\documentclass[preprint,superscriptaddress,aps]{revtex4}

\usepackage{graphicx}
\usepackage{dcolumn}
\usepackage{bm}

\begin{document}


\title{Multiple Superconducting Phases in Palladium Deuteride Induced by Nuclear-Spin Isotope Effect} 

\author{Ryoma Kato}
\affiliation{Department of Applied Quantum Physics, Kyushu University, Motooka, Fukuoka 819-0395, Japan}
\author{Ten-ichiro Yoshida}
\affiliation{Department of Applied Quantum Physics, Kyushu University, Motooka, Fukuoka 819-0395, Japan}
\author{Riku Iimori}
\affiliation{Department of Physics, Kyushu University, Motooka, Fukuoka 819-0395, Japan}
\author{Masanobu Shiga}
\affiliation{Department of Applied Quantum Physics, Kyushu University, Motooka, Fukuoka 819-0395, Japan}
\author{Yuji Inagaki}
\affiliation{Institute for the Advancement of Higher Education, Okayama University of Science,
Ridaicho, Kita-ku, Okayama 700-0005, Japan}
\author{Takashi Kimura}
\affiliation{Department of Physics, Kyushu University, Motooka, Fukuoka 819-0395, Japan}
\author{Koichiro Ienaga}
\affiliation{Graduate School of Sciences and Technology for Innovation, Yamaguchi University, Ube, Yamaguchi, 755-8611, Japan}
\author{Tatsuya Kawae}
\email{kawaetatsuya@gmail.com, t.kawae.122@m.kyushu-u.ac.jp}
\affiliation{Department of Applied Quantum Physics, Kyushu University, Motooka, Fukuoka 819-0395, Japan}


\begin{abstract}
We study the superconducting properties of high-quality PdD$_{x}$ films. Similar to PdH$_{x}$ films, a large residual resistivity remains after a sharp drop in resistivity at $T_{c1}$ $\sim$1.7 K.  In contrast to PdH$_{x}$, the resistivity exhibits a distinct rectangular-shaped anomaly prior to the second transition to zero resistivity at $T_{c2}$ $\sim$0.6 K. The temperature and  magnetic-field dependences of the residual resistivity demonstrate that this anomaly is intrinsic to PdD$_{x}$ and reveal the existence of multiple phases within the superconducting state. These results indicate that the superconducting phase diagram of PdD$_{x}$ differs significantly from that of PdH$_{x}$, highlighting an important role of nuclear spin  ($I$ $=$ 1 for deuterons and $I$ $=$ 1/2 for protons).  

\end{abstract}


\maketitle{}

\section{Introduction}

The quantum nature of the two isotopes of helium has provided a paradigm for quantum many-body phenomena in condensed matter. The difference of neutron numbers  lead to markedly different properties of liquid  $^{4}$He and  $^{3}$He~\cite{1}. Similar differences may be expected between hydrogen (H), with nuclear spin $I$ $=$ 1/2, and deuterium (D), with $I$ $=$ 1, if a mobile phase of these highly concentrated isotopes can be realized. However, such differences have not been clearly manifested so far, mainly because these isotopes form molecules (H$_{2}$, D$_{2}$) at low temperatures. 

In several transition metals, such as palladium (Pd) and niobium (Nb), H or D atoms occupy interstitial sites at concentrations exceeding those in liquid H$_{2}$ or D$_{2}$~\cite{2}. Therefore, if H or D atoms can move between these sites, the two isotopes may exhibit fundamentally different quantum phenomena.
It is well known that H and D atoms can migrate between interstitial sites via quantum tunneling~\cite{2,3,4}, in which the tunneling proton or deuteron is accompanied by an electron cloud during diffusion~\cite{5,6,7}.  Theoretical studies suggest that this situation changes significantly in the superconducting state. The formation of an energy gap at the Fermi surface reduces the coupling between the H or D nuclei and conduction electrons, thereby enhancing their tunneling probability at low temperatures~\cite{8,9,10,11,12,13}.
PdH$_{x}$ ($x$ $=$ H/Pd), one of the most well-known hydride superconductors, exhibits superconductivity at hydrogen concentrations $x$ $\gtrsim$ 0.72 under ambient pressure~\cite{14,15,16}.  The transition temperature increases with increasing $x$ and reaches $\sim$9 K at  $x$ $=$ 1. In contrast, superconductivity in PdD$_{x}$ emerges above $x$ $\sim$0.67 and reaches $\sim$11 K at $x$ $=$ 1~\cite{15,16}. These features are known as the inverse isotope effect, although its origin remains unclear. 

We previously investigated the superconducting properties of PdH$_{x}$ with a superconducting volume fraction of $\sim$1, where hydrogen atoms are loaded into Pd below $T$ $=$ 200 K~\cite{15,17}. Using this method, high-quality samples with a sharp superconducting transition at $T_{c1}$ are obtained. However, despite the sharp transition, a large residual resistivity persists below $T_{c1}$, which subsequently drops to zero at a second transition temperature $T_{c2}$~\cite{18,19,20}. We attribute these features to proton tunneling and its ordering. The tunneling motion of protons may disrupt the global coherence of superconductivity, resulting in finite resistivity, whereas a reduction in the tunneling probability, possibly due to proton ordering, leads to zero resistivity. These findings suggest that tunneling protons in PdH$_{x}$ behave as a quantum liquid or solid, implying that the superconducting properties of PdD$_{x}$, goverrned by tunneling deuterons, are fundamentally different.

In this paper, we study the superconducting properties of a high-quality PdD$_{x}$ film. Similar to PdH$_{x}$ films, a large residual resistivity remains after a sharp drop in resistivity at $T_{c1}$ in  PdD$_{x}$. In contrast to PdH$_{x}$, however, the temperature dependence of resistivity exhibits a rectangular-shaped anomaly prior to the second transition to zero resistivity at $T_{c2}$ $\sim$0.6 K. Based on the temperature and  magnetic-field dependences of the residual resistivity, we show that this anomaly reflects an intrinsic property of the PdD$_{x}$ system and reveals the existence of multiple phases in the superconducting state. These results indicate that the superconducting phase diagram differs significantly between PdD$_{x}$ and PdH$_{x}$ films, highlighting an important role of nuclear spin  ($I$ $=$ 1 for deuterons and $I$ $=$ 1/2 for protons).  This system  therefore provides a platform for investigating quantum many-body phenomena arising from highly concentrated interstitial particles.

\section{Experimental}
The electrical resistivity was measured in a Pd film with a thickness of 100 nm, which is considerably larger than the coherence length of superconducting PdH$_{x}$ and PdD$_{x}$~\cite{15,17}. The sample structure is shown in Fig. 1.  The Pd film was deposited on Au electrodes so that D atoms can be absorbed without obstruction. Au electrodes with a thickness of 20 nm and a width of 50 $\mu$m were fabricated on a  SiO$_{2}$/Si substrate by electron-beam evaporation. To adhere the Au electrodes to the substrate firmly, a 5-nm Ti layer was deposited below the Au electrodes. Subsequently, a Pd film with a thickness of 100 nm and a width of 100 $\mu$m was deposited on the substrate by DC magnetron sputtering at room temperature under a base pressure of 5.5×10$^{-6}$ Pa (see Appendix A). The distance between the two voltage electrodes was 300 $\mu$m. All electrical connections were made using Cu leads, and no superconducting materials were present in the measurement circuit.

Resistivity measurements were conducted in the same manner as in our previous measurements of PdH$_{x}$ films~\cite{19,20}. The Pd film was mounted on a Cu plate installed in a homemade $^{3}$He insert attached to a Quantum Design MPMS SQUID magnetometer~\cite{21}. The resistance was measured using a standard four-wire method using an LR700 AC resistance bridge down to a temperature of $\sim$0.6 K, from which the resistivity was estimated. A RuO$_{2}$ thermometer was attached to the Cu plate. A magnetic field was applied parallel to the film plane and along the current direction. Note that the field deviates from the parallel direction due to deformation of the thin Cu sample holder during mounting . 
  
Deuterium absorption was carried out at temperatures below $T$ = 180 K. As reported previously, exposing metallic Pd to an H$_{2}$ pressure of 0.1 MPa at $T$ = 200 K results in a hydrogen concentration of $x$ $\sim$ 0.8, exceeding the critical concentration for superconductivity in PdH$_{x}$. In the present study,  PdD$_{x}$ films were prepared by low-temperature D absorption under a pressure of $\sim$0.25 MPa, while varying the absorption temperature. The superconducting transition was monitored by in-situ resistivity measurements down to $T$ $\sim$ 0.6 K. 

The Pd film structures and measurement procedures were the same as those used in our previous studies of PdH$_{x}$ films~\cite{20}, with the Pd films fabricated under the same conditions, enabling direct comparison between the two systems.

\begin{figure}
\centering
\includegraphics[width=5.5cm]{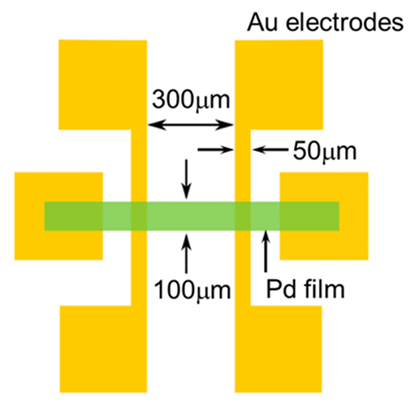}
\caption{(Color online ) A schematic illustration of a Pd sample film. A Pd film with a thickness of 100 nm and a width of 100  $\mu$m is deposited on the substrate with Au electrodes, which have a thickness of 20 nm and a width of 50  $\mu$m. The electrodes are prepared on a SiO$_{2}$/Si substrate with a 5-nm Ti layer deposited below the Au electrodes.  The distance between the two voltage electrodes is 300  $\mu$m.
}
\label{f1}
\end{figure}

\begin{figure}
\centering
\includegraphics[width=8cm]{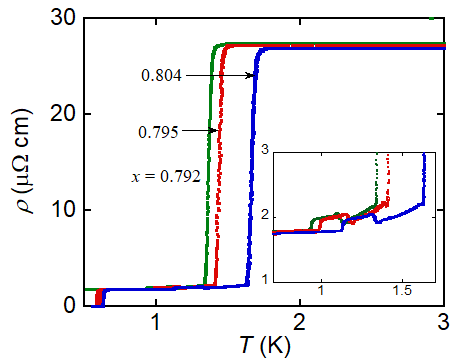}
\caption{(Color online)Temperature dependence of the resistivity in PdD$_{x}$ film for  $x$ $=$ 0.792, 0.795, and 0.804. The inset shows the low-temperature behavior. A rectangular-shaped anomaly appears in the residual-resistivity region. For instance, $T_{RL}$ = 1.14 K and $T_{RH}$ = 1.36 K, as indicated by arrows, in  PdD$_{0.804}$. 
}
\label{f1}
\end{figure}

\section{Results}
A rectangular-shaped anomaly appears in the residual-resistivity region of PdD$_{x}$ after the superconducting transition, whereas no such feature is observed in PdH$_{x}$, as reported previously~\cite{20}. To examine whether this anomaly is intrinsic, we measured the temperature dependence of the resistivity for different D concentrations.  Figure 2 shows the temperature dependence of the resistivity, $\rho$, below 5 K for $x$ $=$ 0.792, 0.795, and 0.804.  First, the resistivity was measured for the D absorption at  $T$ $=$ 170 K. Above the superconducting transition temperature, the resistivity is nearly constant, whereas it drops sharply at $T_{c1}$ $=$ 1.47 K, from which the deuterium concentration is estimated to be $x$ $=$ 0.795~\cite{16}. As in  PdH$_{x}$, a large residual resistivity remains below $T_{c1}$, followed by a second transition to zero resistivity at $T_{c2}$ $\sim$ 0.6 K. In addition, a rectangular-shaped anomaly appears between $T_{RL}$ and  $T_{RH}$ (inset of Fig. 2). Notably, the anomaly at $T_{RH}$ is accompanied by an increase in resistance, in contrast to the sharp resistive drop at $T_{c1}$, suggesting that this anomaly is unlikely to be associated with the development of a bulk superconducting state. Moreover, these features  are independent of the measurement current.

\begin{figure*}[!t]
\centering
\includegraphics[width=16cm]{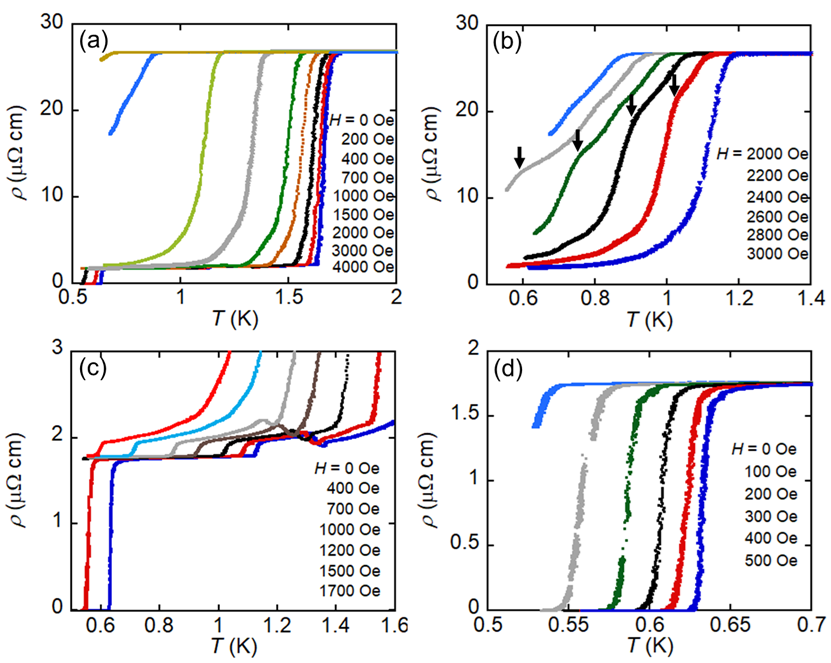}
\caption{ (Color online) (a)  The overall features for magnetic field dependence of the resistivity in the PdD$_{0.804}$ film. The transition curve around $T_{c1}$ is broadened with increasing the magnetic field. (b) Magnetic field dependence of the resistivity between 2000 Oe and 3000 Oe, where magnetic fields are shown in increments of 200 Oe.  A clear bend is visible at  $T_{b}$ in the temperature dependence of the resistivity, as indicated by arrows, between $H$ $=$ 2200 Oe and 2800 Oe. (c) Magnetic field dependence of the rectangular-shaped anomaly. (d)  Magnetic field dependence of the resistivity around $T_{c2}$. A sharp drop in resistivity can be observed up to $H$ $=$ 400 Oe, whereas  zero resistivity is no longer observed at 500 Oe.
}
\label{f1}
\end{figure*}

After the resistivity measurements of the PdD$_{0.795}$  film, the temperature was increased to 160 K, which is lower than the initial absorption temperature.  D$_{2}$ gas was then reintroduced, and the film was reloaded with deuterium. After reabsorption, the low-temperature resistivity was measured again. The superconducting transition temperature increases to $T_{c1}$ $=$ 1.71 K, corresponding to the deuterium concentration of $x$ $=$ 0.804. The second transition also shifts to $T_{c2}$ $=$ 0.64 K. Subsequently, D atoms were desorbed from the film at $T$ $=$ 180 K, resulting in a decrease of $T_{c1}$ to 1.39 K. The corresponding data are plotted for $x$ $=$ 0.792~\cite{16}, where the second transition is not visible. Notably, the rectangular-shaped anomaly shifts systematically during the D absorption and desorption processes without a change in its shape. This demonstrates that the anomaly does not originate from D inhomogeneity but instead reflects an intrinsic property of the PdD$_{x}$ film. A new state, characterized by the rectangular-shaped anomaly, thus emerges in the superconducting PdD$_{x}$ film.

Figure 3(a) illustrates the magnetic-field dependence of the resistivity ($\rho$ – $T$ curves) below $T$ $=$ 2 K in the PdD$_{0.804}$ film. As the magnetic field increases, the superconducting transition temperature $T_{c1}$ is suppressed, accompanied by a broadening of the transition curve, similar to the magnetic-field dependence of $T_{c1}$ in PdH$_{x}$. Notably, a closer examination of the field dependence of the resistivity reveals a distinct difference from that of PdH$_{x}$. Above $H$ $=$ 2200 Oe, a clear bend, referred to as $T_{b}$, appears in the $\rho$ – $T$  curves, as indicated by arrows in Fig. 3(b). The position of this bend shifts to lower temperatures  with increasing magnetic field and is no longer detectable above $H$ $=$ 3000 Oe because it falls below the minimum measurement temperature of $T$ $\sim$ 0.6 K.

The magnetic-field dependence of the resistivity in the low-field region of the  PdD$_{0.804}$ film is presented in Fig. 3(c). At $H$ $=$ 0 Oe, after the superconducting transition, the resistivity decreases gradually with decreasing temperature. A rectangular-shaped anomaly appears between $T_{RL}$ $\sim$ 1.1 K and $T_{RH}$ $\sim$ 1.4 K in the $\rho$ – $T$ curve. As the magnetic field increases, the  anomaly shifts to lower temperatures while its width increases. Above $H$ $=$ 1300 Oe, the upturn at  $T_{RH}$ is no longer visible due to overlap with the broadened tail of the superconducting transition associated with the suppression of $T_{c1}$. In contrast, the resistivity drop at  $T_{RL}$ can be observed up to $H$ $=$ 1700 Oe, where the characteristic shape of the anomaly is preserved. Above $H$ $=$ 1800 Oe, the resistivity drop is no longer visible because $T_{RL}$ shifts below the minimum measurement temperature. 

The $\rho$ – $T$ curves around $T_{c2}$ are shown in Fig. 3(d). A sharp drop in resistivity is observed up to $H$ $=$ 400 Oe, which corresponds to the maximum field at which zero resistivity is detected. At $H$ $=$ 500 Oe, although a resistivity drop is still visible, zero resistivity is no longer observed. At $H$ $=$ 0 Oe, the resistivity starts to deviate at $T$ $\sim$ 0.68 K and decreases to zero at  $T$ $\sim$ 0.62 K, whereas at $H$ $=$ 400 Oe, these temperatures shift to  $T$ $\sim$ 0.59 K and $\sim$0.54 K, respectively. The absence of broadening of the transition curve is consistent with that observed in PdH$_{x}$ and  may therefore  be interpreted as originating from the ordering of tunneling particles~\cite{20}.

\begin{figure}
\centering
\includegraphics[width=8cm]{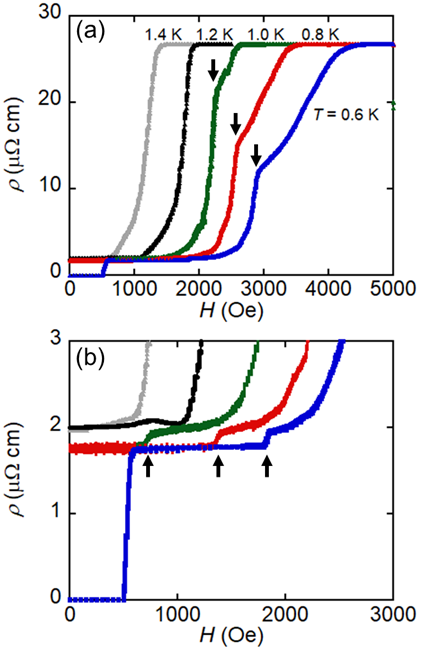}
\caption{ (Color online) (a) An overview on the magnetic-field dependence of the resistivity at $T$ $\sim$0.6 K, 0.8 K, 1.0 K, 1.2 K, and 1.4 K  in the PdD$_{0.804}$ film.  A clear bend appears at $H_{AH}$ in the magnetic field dependence of the resistivity below the upper critical field, $H_{UC}$, as indicated by arrows. (b) The low-field region is extended. The resistivity  at $T$ $\sim$0.6 K, 0.8 K, and 1.0 K increases sharply at $H_{AL}$, as indicated by arrows, as the magnetic field increases.
}
\label{f1}
\end{figure}

Next, we present the magnetic-field dependence of the resistivity at constant temperature ($\rho$ – $H$ curves) in the PdD$_{0.804}$ film. Figure 4(a) shows an overview of the $\rho$ – $H$ curves at representative temperatures, whereas Fig. 4(b) highlights the low-field region. Several anomalies are clearly observed in the $\rho$ – $H$ curves below $T$ $\sim$ 1.0 K. At $T$ $\sim$0.6 K, as the magnetic field increases, the resistivity rises steeply at $H$ $\sim$500 Oe and reaches a nearly constant value of $\sim$1.76 $\mu$$\Omega$cm at $H_{LC}$ $\sim$610 Oe, which is consistent with the residual resistivity below the rectangular-shaped anomaly. Subsequently, the resistivity remains nearly constant and then exhibits another sudden increase  at $H_{AL}$ $\sim$1800 Oe. Notably, the magnitude of this jump is $\sim$0.17 $\mu$$\Omega$cm, which is in good agreement with the resistivity drop at $T_{RL}$ associated with the rectangular-shaped anomaly. These observations suggest that the anomaly can be induced by a magnetic field, indicating a magnetic origin.

Thereafter, the resistivity increases rapidly up to $H$ $\sim$2900 Oe, above which the increase becomes more gradual. This field is defined as $H_{AH}$. Finally, the resistivity reaches that of the normal state at  $H$  $\sim$4500Oe, corresponding to the upper critical field $H_{UC}$. This value is much higher than $H_{UC}$ $ \sim$3300 Oe at $T$ $\sim$0.6 K in the PdH$_{0.870}$ film, even though $T_{c1}$ $=$ 1.71 K in the PdD$_{0.804}$ film is significantly lower than $T_{c1}$ $=$ 2.31 K in the PdH$_{0.870}$ film. The sample setup was identical for both measurements. In addition, the anomalies at $H_{AL}$ and $H_{AH}$ are not observed in the $\rho$ – $H$ curves of PdH$_{x}$ films, in which the resistivity begins to increase steeply at fields slightly below $H_{UC}$. At $T$ $\sim$0.8 K and $\sim$1.0 K, similar features are observed, where the anomalies at $H_{AL}$, $H_{AH}$, and $H_{UC}$ shift to lower magnetic fields, whereas the jump at $H_{LC}$ is no longer detectable. Above $T$ $\sim$1.2 K, the anomalies at $H_{LC}$, $H_{AL}$, and $H_{AH}$ are no longer observed, and only the anomaly associated with the upper critical field remains.

Based on the anomalies observed in the  $\rho$ – $T$ and  $\rho$ – $H$ measurements, we construct the $T$ - $H$ phase diagram shown in Fig. 5(a). The anomalies obtained from the two measurements are in good agreement, indicating that those observed in the resistivity originate from changes in the superconducting properties of the PdD$_{0.804}$ film. In other words, new phases emerge between $T_{c1}$ and $T_{c2}$ within the superconducting state, demonstrating the presence of multiple phases. Importantly, similar behavior is also observed in the PdD$_{0.792}$ film (see Appendix B), confirming that the multiple-phase structure is a robust feature of PdD$_{x}$. In contrast, the temperature dependence of the resistivity in the PdH$_{0.870}$ film exhibits only two transitions at $T_{c1}$ and $T_{c2}$, as shown in the phase diagram in Fig. 5(b)~\cite{20}. The two phase diagrams are clearly different, despite the hydride and deuteride samples being prepared by the same procedure.

\section{Discussion}
We discuss the possible origin of the emergence of multiple phases. One possibility is the phase separation due to spatial variations in the deuterium concentration within the film.  However,  the emergence of such a large number of phases is unlikely to be explained solely by inhomogeneity in the PdD$_{x}$ film. Similarly, a superconducting contribution from an Au–Pd alloy can be ruled out~\cite{22}, because such multiple phases are not observed in PdH$_{x}$ films prepared under identical conditions. Moreover, even excluding the phases below $T_{c2}$, at least three distinct phase transitions are identified in zero magnetic field. In addition, a Josephson network at grain boundaries cannot account for the  emergence of multiple phases.

\begin{figure}
\centering
\includegraphics[width=8cm]{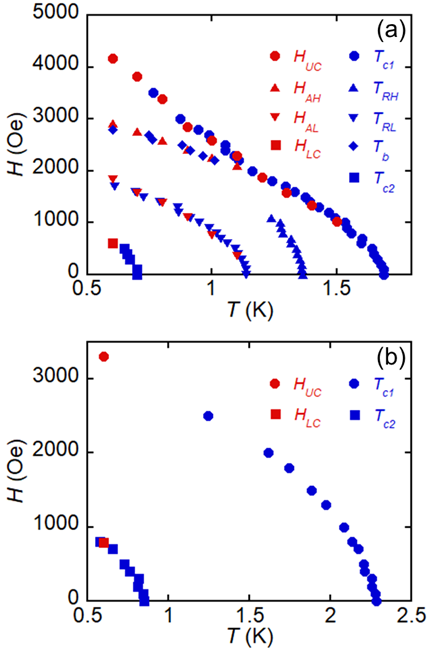}
\caption{(Color online) The $T$ - $H$ phase diagram of the PdD$_{0.804}$ film (a)  and  PdH$_{0.870}$ film (b)~\cite{20}. The blue and red plots are obtained based on the $\rho$ – $T$ and  $\rho$ – $H$ measurements, respectively. 
}
\label{f1}
\end{figure}

From these observations, we conclude that the multiple-phase structure between $T_{c1}$ and $T_{c2}$ reflects intrinsic superconducting properties of PdD$_{x}$. This indicates that the isotope effect between PdH$_{x}$ and PdD$_{x}$ manifests as a qualitative difference in their superconducting phase diagrams.  The results further suggest that the nuclear spin moment ($I$ $=$ 1) of deuterons plays an important role in the emergence of the multiple-phase diagram in PdD$_{x}$. In contrast, the mass difference, which affects the tunneling probability, is unlikely to be the primary origin. If it were dominant, a more systematic variation of the phase diagram with isotope mass would be expected, which is not observed.

Next, we consider why multiple phases emerge only in the superconducting state of PdD$_{x}$. As shown in Figs. 3 and 4, the $\rho$ – $T$ and $\rho$ – $H$ curves exhibit clear kinks in their temperature and magnetic-field dependences, indicating that the phase boundaries arise from phase transitions. One possible origin is magnetic ordering of nuclear moments, which would modify the scattering between the nuclear spins and Cooper pairs. In this case, the ordering temperature of deuterons ($I$ $=$ 1) is expected to be higher than that of protons ($I$ $=$ 1/2), and the additional degrees of freedom associated with $I$ $=$ 1 could allow the emergence of multiple phases. However, this scenario is unlikely, because nuclear magnetic ordering temperatures in metals are typically below $\sim$$\mu$K~\cite{23}. Similarly, a scenario based on the formation of Yu–Shiba–Rusinov states is unlikely to explain the multiple-phase diagram, because the relevant energy scale, determined by the coupling to nuclear moments, is extremely small~\cite{24}.

We therefore attribute the multiple superconducting phases to variations of the superconducting order parameter across phase boundaries, implying the emergence of non-$s$-wave pairing~\cite{25}. In hydride superconductors, pairing is generally considered to be $s$-wave, driven by strong electron–phonon coupling due to large-amplitude vibrations of light hydrogen atoms~\cite{26}. However,  PdD$_{x}$ exhibits anomalous features, such as the inverse isotope effect and a negative pressure dependence of the superconducting transition temperature~\cite{27}, which are difficult to reconcile with conventional $s$-wave superconductivity. Theoretically, strong coupling between electrons and large-amplitude hydrogen/deuteron vibrations can induce anharmonicity in the vibrational potential, which has been proposed as the origin of the inverse isotope effect~\cite{28,29,30,31,32}. Such anharmonicity can give rise to anisotropic electron–phonon interactions. Moreover, collective motion of deuterons with bosonic character can enhance direction-dependent tunneling processes, leading to anisotropic coupling to electrons and thereby promoting non-$s$-wave pairing in PdD$_{x}$. In contrast, for PdH$_{x}$, the fermionic nature of protons introduces exchange antisymmetry in their wave function, which tends to suppress collective motion and effectively averages out anisotropic scattering. As a result, conventional $s$-wave pairing is more likely to be stabilized in PdH$_{x}$. Taken together, these considerations offer a plausible explanation for the observed differences between PdD$_{x}$ and PdH$_{x}$, although further investigation is required.

\section{Conclusion}
We have studied the superconducting properties of high-quality PdD$_{x}$ films prepared at low temperatures. The PdD$_{x}$ films exhibit a large residual resistivity after a sharp drop at $T_{c1}$ in resistivity, similar to PdH$_{x}$. In contrast, a distinct rectangular-shaped anomaly appears prior to the second transition to zero resistivity at $T_{c2}$. Based on the temperature and magnetic-field dependences of the resistivity, we demonstrate the existence of multiple phases in the superconducting state of PdD$_{x}$. The results reveal a clear difference in the superconducting phase diagrams between  PdD$_{x}$ and PdH$_{x}$, highlighting the important role of the nuclear-spin isotope effect ($I$ $=$ 1 for deuterons and $I$ $=$ 1/2 for protons) in tunneling dynamics within the superconducting state. This system therefore  provides a new platform for investigating quantum statistics in a dense ensemble of quantum particles.

{\bf Acknowledgments}
We specially thank Profs. Masashige Matsumoto, Keiya Shirahama, Isao Maruyama, and Youichi Yanase for fruitful discussion. We also thank Professor Osamu Yamamuro for providing information about the low-temperature H absorption into Pd. We also thank Mrs. Tadahiko Hasuo and Kyohei Yamaguchi for their technical help. This work was supported by JSPS KAKENHI Grant Numbers JP23K17763 and JP21H01605, Nippon Sheet Glass Foundation for Materials Science and Engineering, and Murata Science Foundation.

 {\bf Data availability} 
The data that support the findings of this study are available from the corresponding author upon reasonable request.

\appendix

\renewcommand{\thefigure}{\Alph{section}-\arabic{figure}}
\makeatletter
\@addtoreset{figure}{section}
\makeatother

\section{Fabrication and characterization of Pd films}

The deposition conditions of the Pd films are summarized in Table A1. The films were deposited at room temperature by DC magnetron sputtering under an Ar atmosphere.
Figure A1 shows the X-ray diffraction pattern of a reference Pd film prepared simultaneously under identical conditions  as those used for the Pd film in the present measurements. The diffraction data indicate a polycrystalline structure with a pronounced (111) preferred orientation. While several peaks expected for randomly oriented Pd powder (e.g., (220)) are strongly suppressed, the Pd(111) reflection is clearly observed, indicating a strong preferred orientation rather than a randomly oriented polycrystalline structure. Similar diffraction features were consistently observed for films prepared under identical conditions.

\begin{table*}[t]
\caption{Deposition conditions of Pd films}
\label{t1}
\begin{center}
\begin{tabular}{ll}

\hline
Deposition method & DC magnetron sputtering \\
Current & 0.17 A (Voltage: 399 V) \\
Base pressure & 5.5 $\times$ 10$^{-6}$ Pa \\
Ar pressure & 2.45 $\times$ 10$^{-1}$Pa (Ar flow: 10 sccm) \\
Deposition rate & 63.9 nm / 2 min (approx. 32 nm/min) \\
Substrate temperature & Room temperature (no heating during deposition) \\

\hline
\end{tabular}
\end{center}
\end{table*}

\begin{figure}
\centering
\includegraphics[width=9cm]{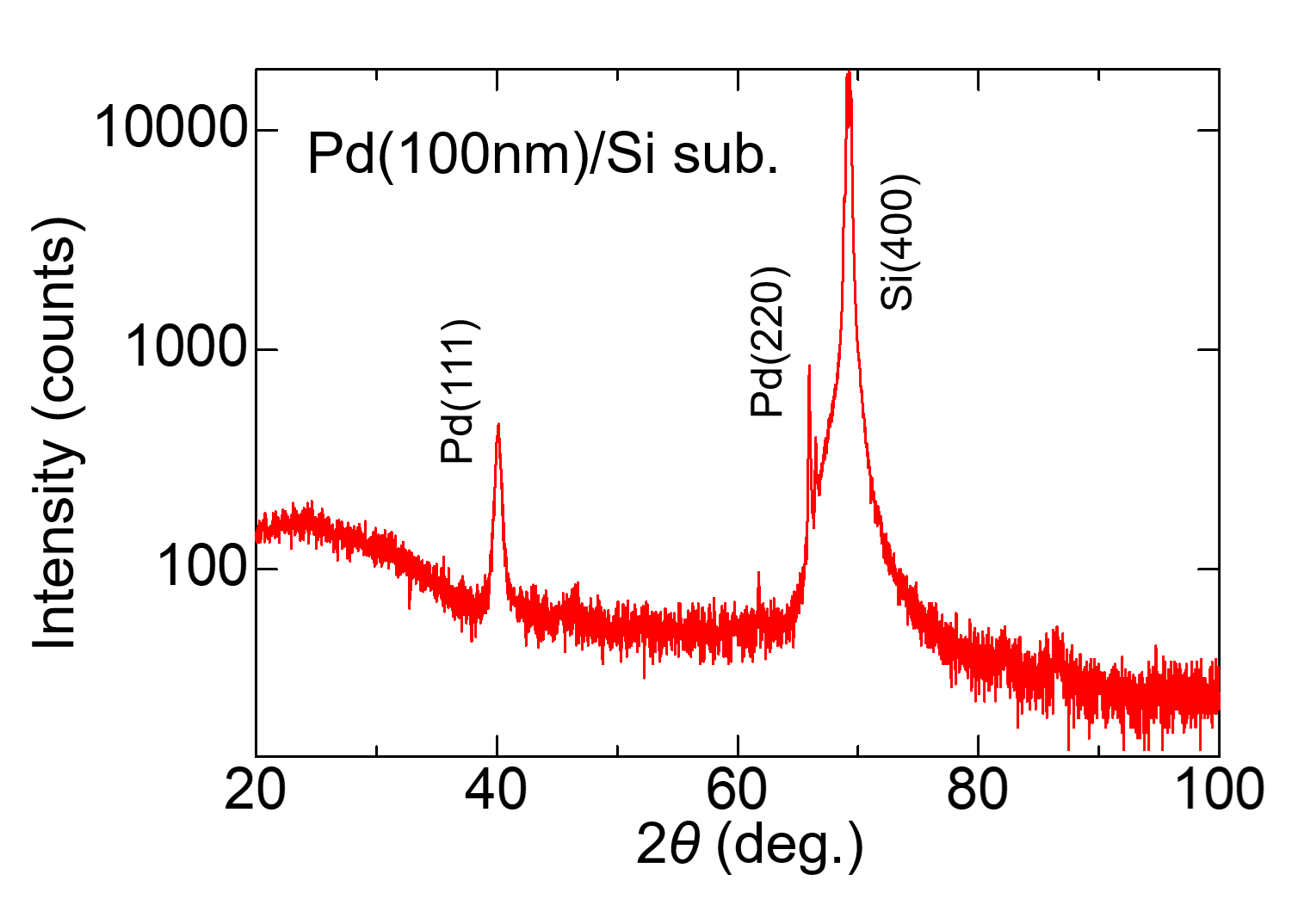}
\caption{ (Color online) X-ray diffraction pattern of a reference Pd film prepared simultaneously under identical conditions as the Pd film used in the present measurements.
}
\label{f1}
\end{figure}

\vspace{\baselineskip}
\section{Magnetic field dependence of electrical resistivity in PdD$_{0.792}$ }

\begin{figure}
\centering
\includegraphics[width=8cm]{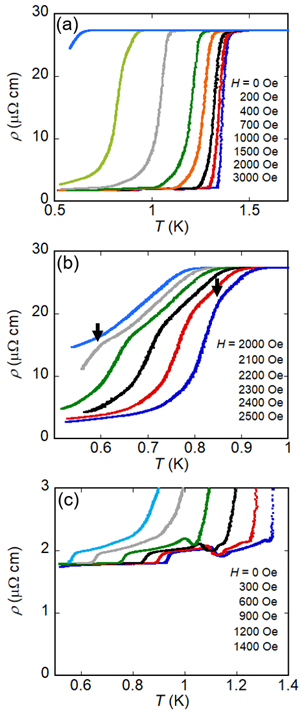}
\caption{ (Color online) (a)  The overall features for magnetic field dependence of the resistivity in the PdD$_{0.792}$ film. The transition curve around $T_{c1}$ is broadened with increasing the magnetic field. (b) Magnetic field dependence of the resistivity between 2000 Oe and 2500 Oe, where magnetic fields are shown in increments of 100 Oe.  A clear bend is visible at  $T_{b}$ in the temperature dependence of the resistivity, as indicated by arrows, between $H$ $=$ 2000 Oe and 2400 Oe. (c) Magnetic field dependence of the rectangular-shaped anomaly.
}
\label{f1}
\end{figure}

\begin{figure}
\centering
\includegraphics[width=8cm]{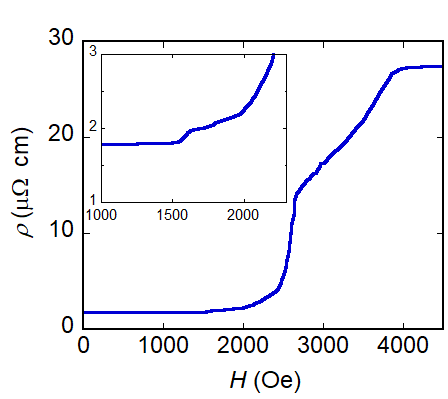}
\caption{ (Color online) An overview on the magnetic-field dependence of the resistivity at $T$ $\sim$0.5 K,  in the PdD$_{0.792}$ film.  A clear bend appears at $H_{AH}$  $\sim$2650 Oe in the magnetic field dependence of the resistivity below the upper critical field, $H_{UC}$. The low-field region is extended in the inset. The resistivity increases sharply at $H_{AL}$ $\sim$ 1610 Oe, as the magnetic field increases.
}
\label{f1}
\end{figure}

\begin{figure}
\centering
\includegraphics[width=8cm]{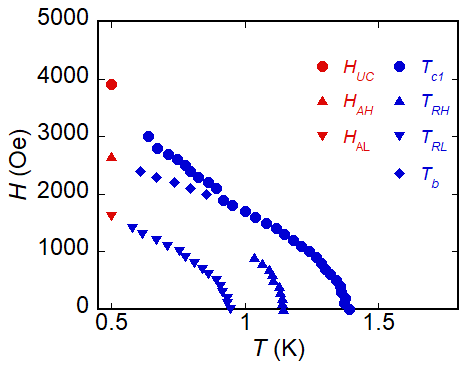}
\caption{(Color online) The $T$ - $H$ phase diagram of the PdD$_{0.792}$ film. The blue and red plots are obtained based on the $\rho$ – $T$ and  $\rho$ – $H$ measurements, respectively. 
}
\label{f1}
\end{figure}

Figures B1(a)-B1(c) show the magnetic-field dependence of the resistivity ($\rho$ – $T$ curves) in the PdD$_{0.792}$ film. As the magnetic field increases, the superconducting transition temperature $T_{c1}$ is suppressed, accompanied by a broadening of the transition curve, similar to the behavior observed in the PdD$_{0.804}$ film. However, the upper critical field in the PdD$_{0.792}$ film is lower than that in the PdD$_{0.804}$ film. Note that the magnetic-field directionis is identical for both films, since the deuterium concentration was varied by changing the absorption temperature within the same experimental setup. Figure B1(b) illustrates the magnetic-field dependence of $T_{b}$, defined as the bend in the $\rho$ – $T$  curve. The bend appears at $H$ $=$ 2000 Oe, whereas it is observed above $H$ $=$ 2200 Oe in the PdD$_{0.804}$. The position of  $T_{b}$ shifts to lower temperatures with increasing magnetic field and is no longer detectable above $H$ $=$ 2500 Oe. These characteric fields are also lower than those in PdD$_{0.804}$.

Figure B1(c) shows  the low-field behavior of the resistivity. At $H$ $=$ 0 Oe, a rectangular-shaped anomaly appears between $T_{RL}$ $\sim$ 0.9 K and  $T_{RH}$ $\sim$ 1.1 K. As the magnetic field increases, the anomaly shifts to lower temperatures while its width increases. Above $H$ $=$ 1100 Oe, the upturn at $T_{RH}$ is no longer visible due to overlap with the broadened superconducting transition. In contrast, the resistivity drop at  $T_{RL}$ persists up to $H$ $=$ 1400 Oe, retaining its characteristic shape. Above $H$ $=$ 1500 Oe, the drop disappears because  $T_{RL}$ shifts below the minimum measurement temperature. No resistivity drop at $T_{c2}$ is observed even at $H$ $=$ 0 Oe.

 Figure B2 presents the magnetic-field dependence of the resistivity at constant temperature ($T$ $\sim$ 0.5 K). The overall behavior is consistent with that observed in the PdD$_{0.804}$ film. Because no transition to zero resistivity occurs, the resistivity remains nearly constant ($\sim$1.78 $\mu$$\Omega$cm) at low fields. It then exhibits a step-like increase at $H_{AL}$ $\sim$ 1610 Oe, as shown in the inset. The magnitude of this jump agrees well with the resistivity drop at $T_{RL}$ associated with the rectangular-shaped anomaly. The resistivity subsequently increases rapidly up to $H_{AH}$ $\sim$ 2650 Oe, above which the increase becomes more gradual, and finally reaches the normal state value at $H_{UC}$ $\sim$ 3900 Oe. 

Based on those  anomalies, we construct the $T$ – $H$ phase diagram shown in Fig. B3. As in  PdD$_{0.804}$ [Fig. 5(a)], multiple phases are identified within the superconducting state, although the characteristic temperatures and fields are systematically lower.

\end{document}